\documentclass[10pt]{cernrep}
\usepackage{graphicx}
\usepackage{floatflt}

\newcommand{\Pt}{{P_t}}

\newcommand{\gpj}{~"$\gamma+Jet$"~}

\newcommand{\ptgj}{$~\Pt^{\gamma}$ and $\Pt^{jet}~$}
\newcommand{\ptg}{$\Pt^{\gamma}$}

\suppressfloats[!]

\begin{document}


\thispagestyle{empty}

\vskip-5mm

\begin{center}
{\Large JOINT INSTITUTE FOR NUCLEAR RECEARCH}
\end{center}

\vskip10mm

\begin{flushright}
JINR Preprint \\
E2-2000-254 \\
hep-ex/0011014
\end{flushright}

\vspace*{3cm}

\begin{center}
\noindent
{\Large{\bfseries Jet energy scale setting with \gpj events at LHC \\[0pt]
energies. Selection of events with a clean \gpj topology\\[0pt]
and $\Pt^{\gamma} - \Pt^{Jet}$ disbalance.}}\\[5mm]
{\large D.V.~Bandourin$^{1\,\dag}$, V.F.~Konoplyanikov$^{2\,\ast}$, 
N.B.~Skachkov$^{3\,\dag}$}

\vskip 0mm

{\small
{\it
E-mail: (1) dmv@cv.jinr.ru, (2) kon@cv.jinr.ru, (3) skachkov@cv.jinr.ru}}\\[3mm]
$\dag$ \large \it Laboratory of Nuclear Problems \\
\hspace*{-4mm} $\ast$ \large \it Laboratory of Particle Physics
\end{center}

\vskip 9mm
\begin{center}
\begin{minipage}{150mm}
\centerline{\bf Abstract}
\noindent
It is shown in the paper that $\Pt$ activity limitation (modulus of
the vector sum) of all particle beyond \gpj system $\Pt^{out}$ leads
to the noticeable \ptgj disbalance decreasing. On a simultaneous restriction
of the cluster $\Pt$ and $\Pt^{out}$ from above it is possible to reach
an acceptable balance between \ptgj with a sufficient number of the \gpj
events for the jet energy scale setting and hadron calorimeter calibration
of the CMS detector at LHC. 
\end{minipage}
\end{center}

\newpage

\vskip2cm
\setcounter{page}{1}
\section{INTRODUCTION} 

Here we continue a detailed study of the  \ptgj disbalance and are planning
to show how the calibration accuracy can be improved by simultaneous imposing both
cuts $\Pt^{clust}_{CUT}$ and $\Pt^{out}_{CUT}$ as well as by introduction
of jet isolation requirement.

\section{DETAILS OF ${\bf \Pt^{\gamma}}$ and ${\bf \Pt^{Jet}}$ DISBALANCE DEPENDENCE
ON $\Pt^{clust}_{CUT}$ AND $\Pt^{out}_{CUT}$ PARAMETERS.}

In the previous papers ([1, 2]) we introduced observables
(variables) and discussed what cuts for them may lead to a decrease in the
$\Pt^{\gamma}$ and $\Pt^{Jet}$ disbalance.
Below we concentrate on three of them:
a restriction of cluster $\Pt$ ($\Pt^{clust}_{CUT}$), limitation of
the summed vector $\Pt$ of all particles detectable in the $|\eta|<5$ region
out of the \gpj system ($\Pt^{out}_{CUT}$), the cut for jet isolation
($\epsilon^{jet}$)
\footnote{For the detailed explanation of cuts used here we refer the reader
to papers [1--3].}.

Figs.~1--7 can be considered as an illustration and a complement
to the tables of Appendixes 1--4 of [3].
In Fig.~\ref{fig:j-clu} we show a dependence of the ratio
$(\Pt^{\gamma}\!-\!\Pt^{J})/\Pt^{\gamma}$ on the $\Pt^{clust}_{CUT}$ value
for the case of Selection 1 and two jetfinders LUCELL and UA1 for
two  \ptg ~
intervals $40<\Pt^{\gamma}<50~ GeV/c$ and $300<\Pt^{\gamma}<360~ GeV/c$.
An evident tendency
of balance improvement with decreasing $\Pt^{clust}_{CUT}$
is revealed for all three jetfinding algorithms and both \ptg~($\approx \Pt^{Jet}$)
intervals. It is seen that we can essentially increase the accuracy
by constraining $\Pt^{clust}_{CUT}$ (without $\Pt^{out}$ restriction)
by 5 $GeV/c$.
Thus, for the UA1 algorithm the mean and RMS values of $(\Pt^{\gamma}-\Pt^{J})/\Pt^{\gamma}$ drop from $0.029$ down to $0.021$
and from $0.174$ down to $0.105$, respectively, in the first \ptg ~interval.
For the  LUCELL algorithm the situation after such
a strict cut becomes even better. In Figs.~2--5  the average values for the
$(\Pt^{\gamma}\!-\!\Pt^{J})/\Pt^{\gamma}$ variable and the number of events
for $L_{int}=3\,fb^{-1}$ are displayed for two types
of Selections as a function of $\Pt^{clust}_{CUT}$ for four \ptg ~intervals
and for all three jetfinders. Again, passing to Selection 2 we see that
for all \ptg ~ intervals and for both jetfinders the balance
gradually improves with restricting $\Pt^{clust}_{CUT}$. After limiting
$\Pt$ activity in the ring around the jet (Figs.~4, 5) the disbalance drops to the
$1\%$ level for the $~40<\Pt^{\gamma}<50~ GeV/c~$ interval and
for $\Pt^{clust}_{CUT}=5~ GeV/c$.  The number of events
in this case decreases by a factor of 5 as compared with Selection 1.
It falls
down to 30--50 thousand events at $L_{int}=3\,fb^{-1}$, which seems to be
still quite
sufficient statistics for accurate determination of the jet scale and calibration.
It should be noted that starting from \ptg=$100~ GeV/c$ practically all
events in the Selection 2 sample are comprised inside the $1\%$ accuracy
window. At the same time the number of events decreases by about twofold
with respect to Selection 1.

Up to now we have been studying the influence of the $\Pt^{clust}_{CUT}$
parameter on the balance. Let us see in analogy with Fig.~\ref{fig:j-clu}
 what effect is produced by the $\Pt^{out}_{CUT}$ variation.
If we constrain this variable by 5 $GeV/c$,
keeping $\Pt^{clust}$ weakly restricted by $\Pt^{clust}_{CUT}=30~ GeV/c$
(practically unbound), then, as can be seen from
Fig.~\ref{fig:j-out},
the mean and RMS values of the $(\Pt^{\gamma}\!-\!Pt^J)/\Pt^{\gamma}$
variable in the case of UA1 algorithm decrease
from $3\%$ down to $1.6\%$ and from $17.4\%$
down to $8.8\%$, respectively, for $40<\Pt^{\gamma}<50~ GeV/c$. For LUCELL jetfinder
the $(\Pt^{\gamma}\!-\\!Pt^J)/\Pt^{\gamma}$ value is even less.
At $300<\Pt^{\gamma}<360~ GeV/c$ practically all events have the mean and RMS
values of $(\Pt^{\gamma}\!-\!\Pt^J)/\Pt^{\gamma}$ less than $1.3\%$ and $5\%$,
respectively.

The influence of the $\Pt^{out}_{CUT}$ variation (with the fixed value
$\Pt^{clust}_{CUT}=10 ~GeV/c$) on the
distribution of $(\Pt^{\gamma}\!-\!\Pt^J)/\Pt^{\gamma}$ is shown in
Fig.~\ref{fig:j-out-} for Selection 1. In this case the mean value drops
from $2\%$ to $1.5\%$ for UA1 and to about $1\%$ for the LUCELL algorithm
for $40<\Pt^{\gamma}<50~ GeV/c$ interval. At the same time
RMS changes from $10-11\%$ to $8\%$ level for all algorithms.

\section{SUMMARY}
The new cuts introduced in [1] $\Pt^{clust}_{CUT}$ and $\Pt^{out}_{CUT}$
as well as introduction of a new object of isolated jet are found to be
very efficient tools to improve the calibration accuracy. Their combined
usage for this aim and for the background suppression will be shown in more
details in paper [4].

\section{ ACKNOWLEDGMENTS}                                         
We are greatly thankful to D.~Denegri for having offered this theme to study,
fruitful discussions and permanent support and encouragement.
It is a pleasure for us
to express our recognition for helpful discussions to P.~Aurenche,
M.~Dittmar, M.~Fontannaz, J.Ph.~Guillet, M.L.~Mangano, E.~Pilon,
H.~Rohringer, S.~Tapprogge and J.~Womersley.


\begin{figure}
\hspace{0mm} \includegraphics[width=13cm]{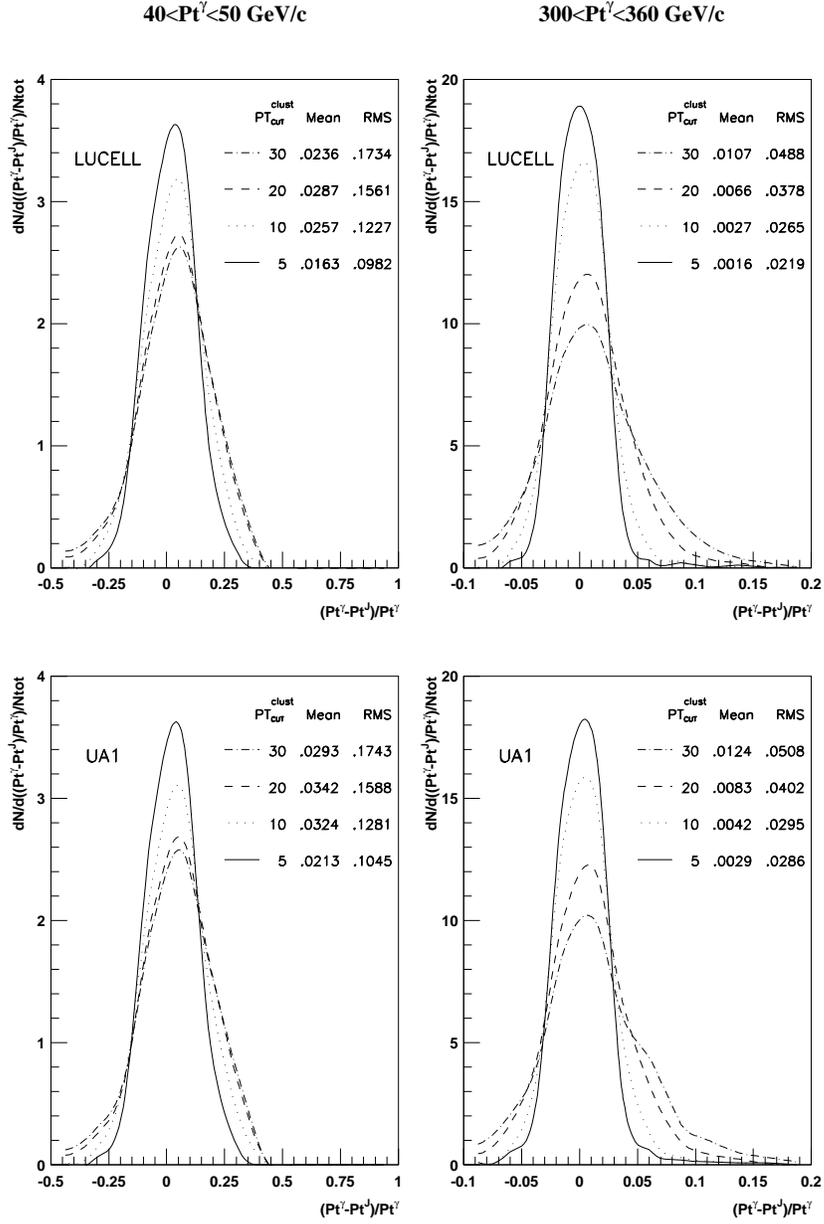}
\caption{A dependence $(\Pt^{\gamma}-\Pt^{J})/\Pt^{\gamma}$ on
$\Pt^{clust}_{CUT}$ for LUCELL and UA1 jetfinding algorithms and two
intervals of \ptg. The mean and RMS of the distributions are displayed on
the figures. $\Pt^{out}$ is not limited. Selection 1}
\label{fig:j-clu}
\end{figure}

\newpage
\begin{flushright}
\parbox[r]{.45\linewidth}{
{\footnotesize Fig.~2 (left):
Selection 1. $\Delta \phi=15^{\circ}$.
Number of events (for $L_{int}=3fb^{-1}$)
dependence on $\Pt^{clust}_{CUT}$ in cases of LUCELL and UA1
jetfinding algorithms. $\Pt^{out}$ is not limited.
\\ \\ \\ \\ \\ \\ \\ \\ \\ \\
\noindent
Fig.~3 (bottom):
Selection 1. $\Delta \phi=15^{\circ}$.
$(\Pt^{\gamma}-\Pt^{J})/\Pt^{\gamma}$
dependence on $\Pt^{clust}_{CUT}$ in cases of UA1 and LUCELL
jetfinding algorithms. $\Pt^{out}$ is not limited.}}
\end{flushright}
\begin{figure}[h]
 \vspace{-9cm}
 \hspace{0.0cm}
\includegraphics[height=8cm,width=7cm]{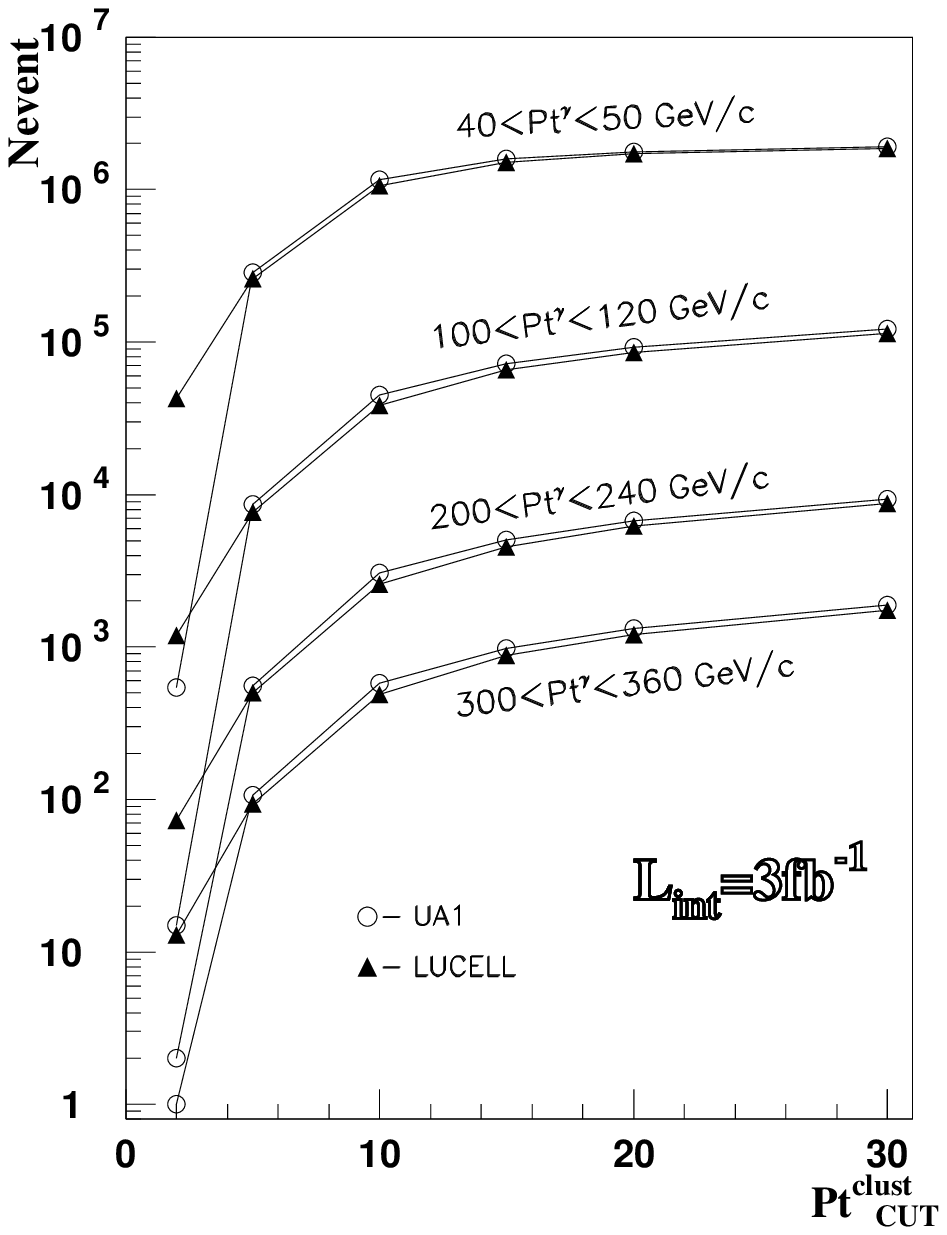}
    \label{fig01}
   \nonumber
  \end{figure}
 \vspace{-1.0cm}
\begin{figure}[h]
 \hspace{2mm}
 \vspace*{-10.0cm}
\includegraphics[height=13cm,width=14cm]{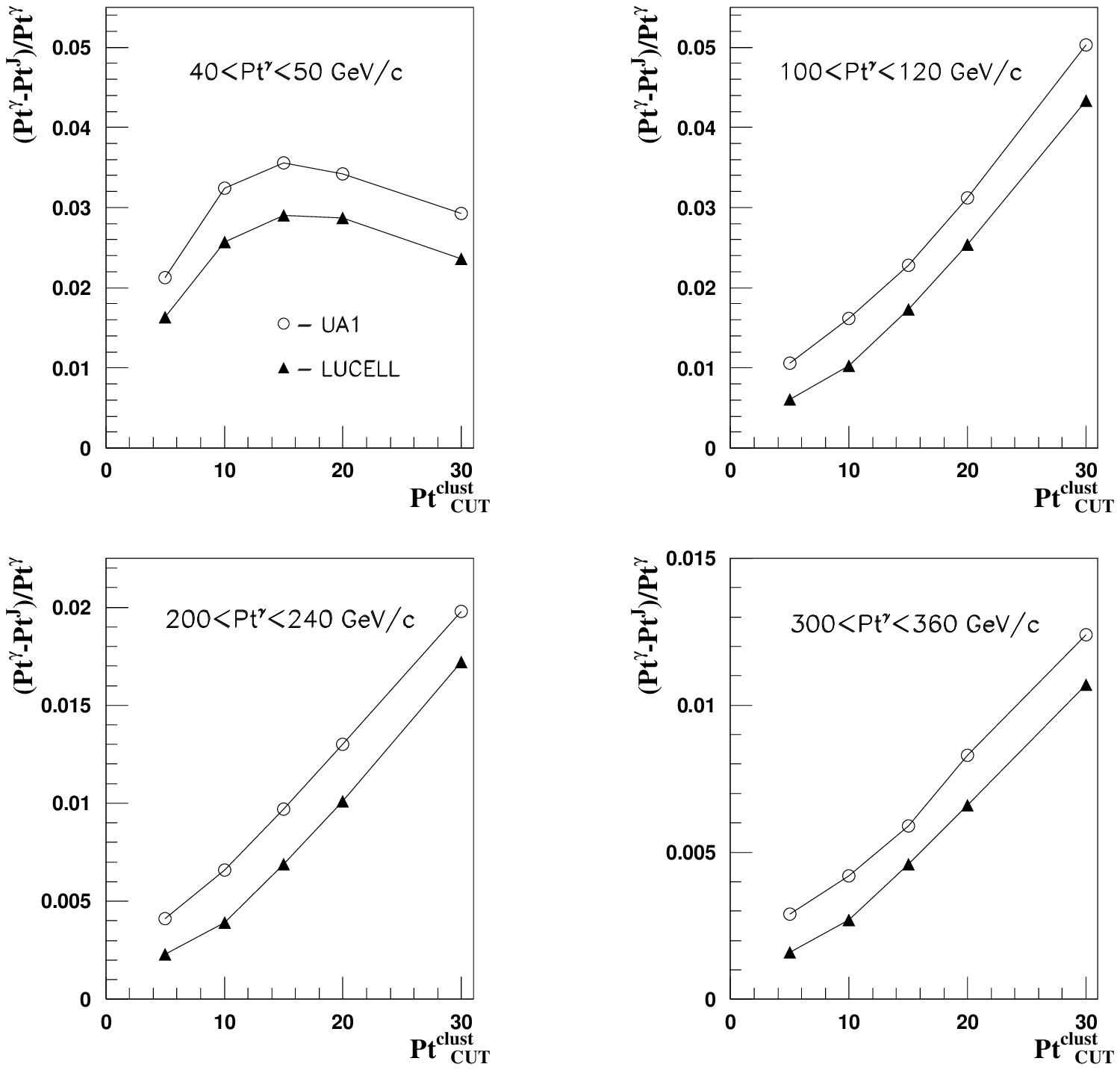}
    \label{fig02}
\nonumber
\end{figure}

\newpage
\begin{flushright}
\parbox[r]{.45\linewidth}{
{\footnotesize Fig.~4 (left):
Selection 2. $\Delta \phi=15^{\circ}$, $\epsilon^{jet}<2\%$.
Number of events (for $L_{int}=3fb^{-1}$)
dependence on $\Pt^{clust}_{CUT}$ in cases of LUCELL and UA1
jetfinding algorithms. $\Pt^{out}$ is not limited.
\\ \\ \\ \\ \\ \\ \\ \\ \\ \\
\noindent
Fig.~5 (bottom):
Selection 2. $\Delta \phi=15^{\circ}$,  $\epsilon^{jet}<2\%$.
$(\Pt^{\gamma}-\Pt^{J})/\Pt^{\gamma}$
dependence on $\Pt^{clust}_{CUT}$ in cases of LUCELL and UA1
jetfinding algorithms. $\Pt^{out}$ is not limited.}}
\end{flushright}
\begin{figure}[h]
 \vspace{-9cm}
 \hspace{0.0cm}
\includegraphics[height=8cm,width=7cm]{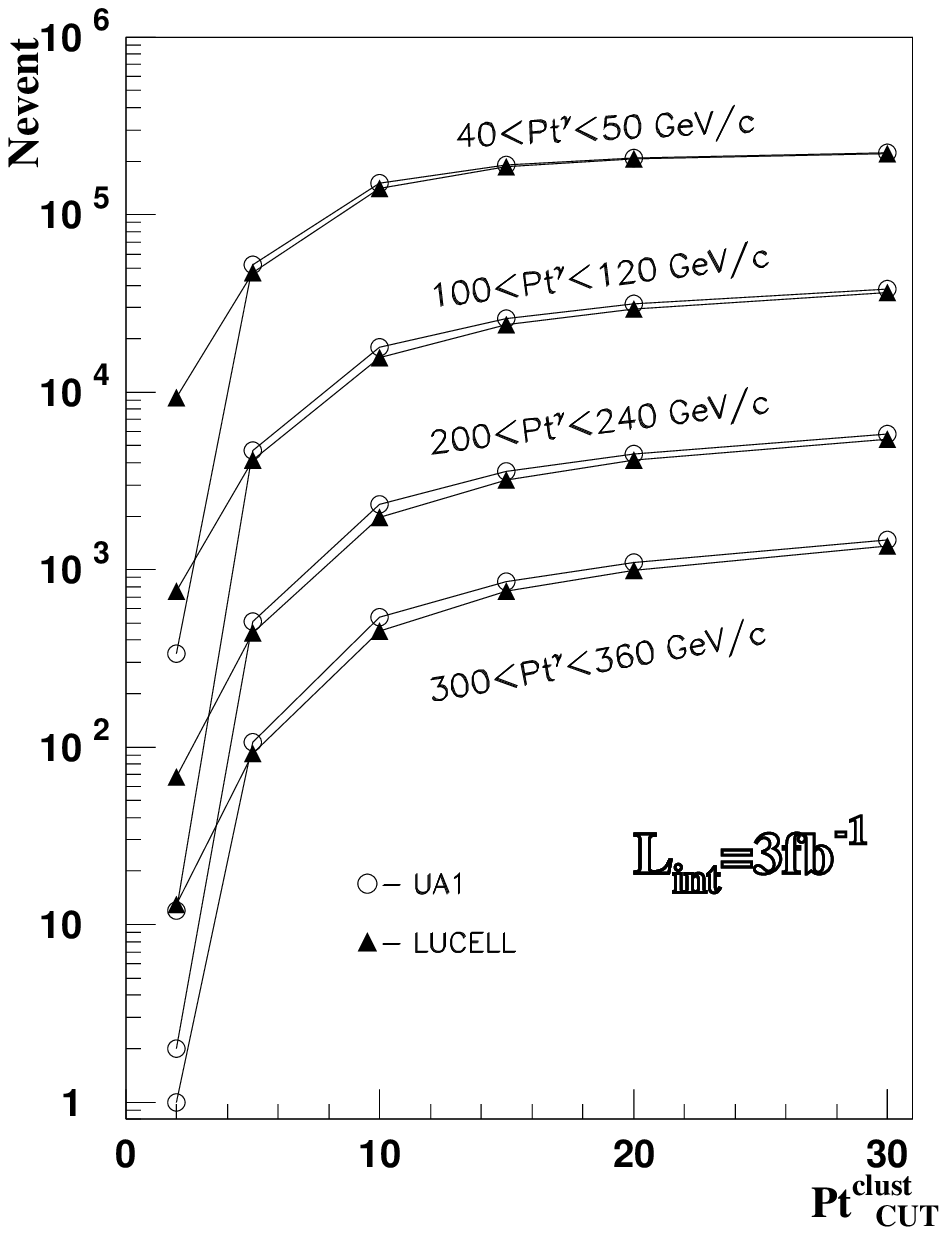}
 \label{fig03}
 \nonumber
 \end{figure}
\vspace{-1.0cm}
\begin{figure}[h]
\hspace{2mm}
\vspace*{-10.0cm}
\includegraphics[height=13cm,width=14cm]{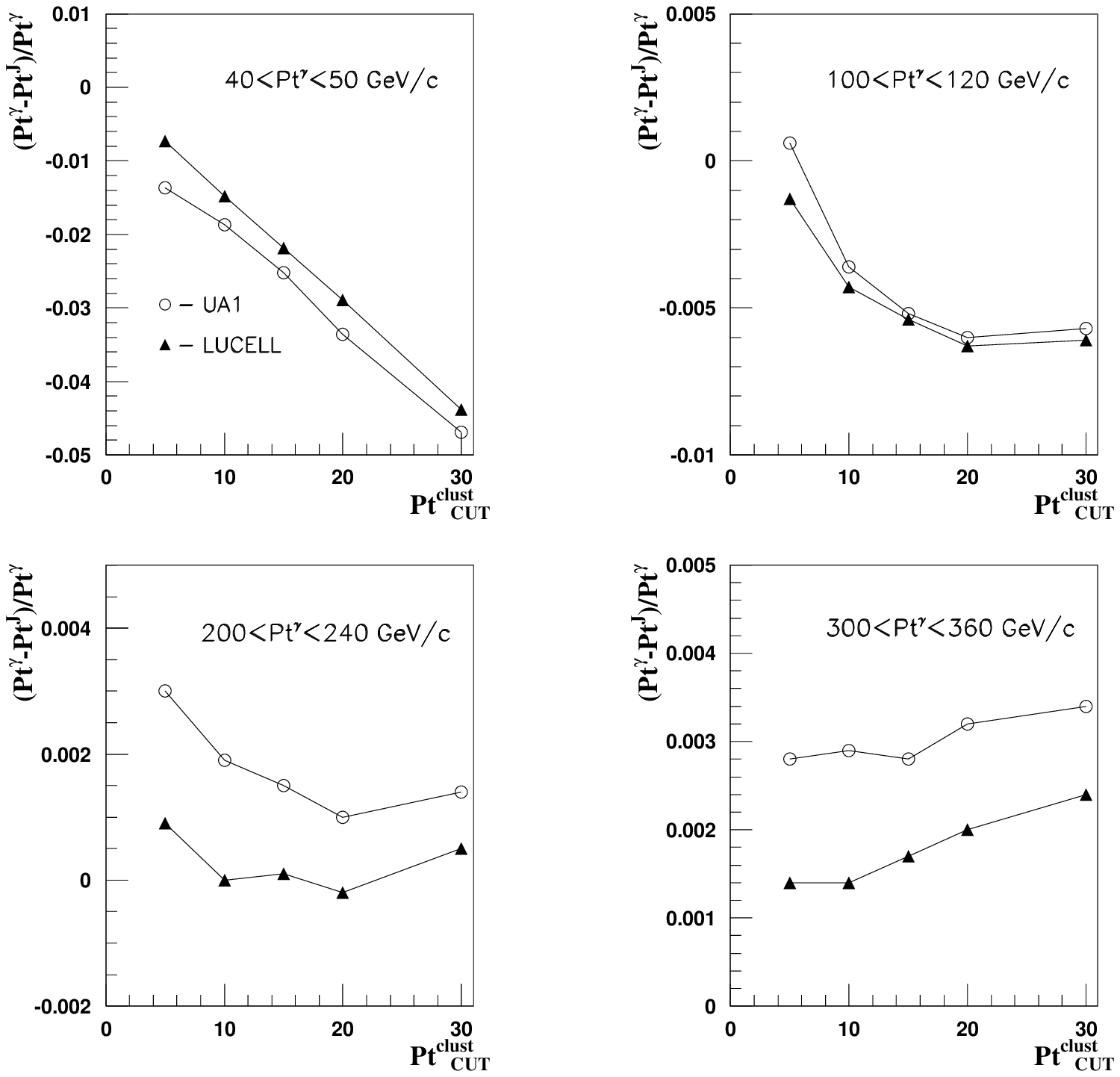}
\label{fig04}
\nonumber
\end{figure}

\newpage

\setcounter{figure}{5}
\begin{figure}
  \hspace{0mm} \includegraphics[width=13cm]{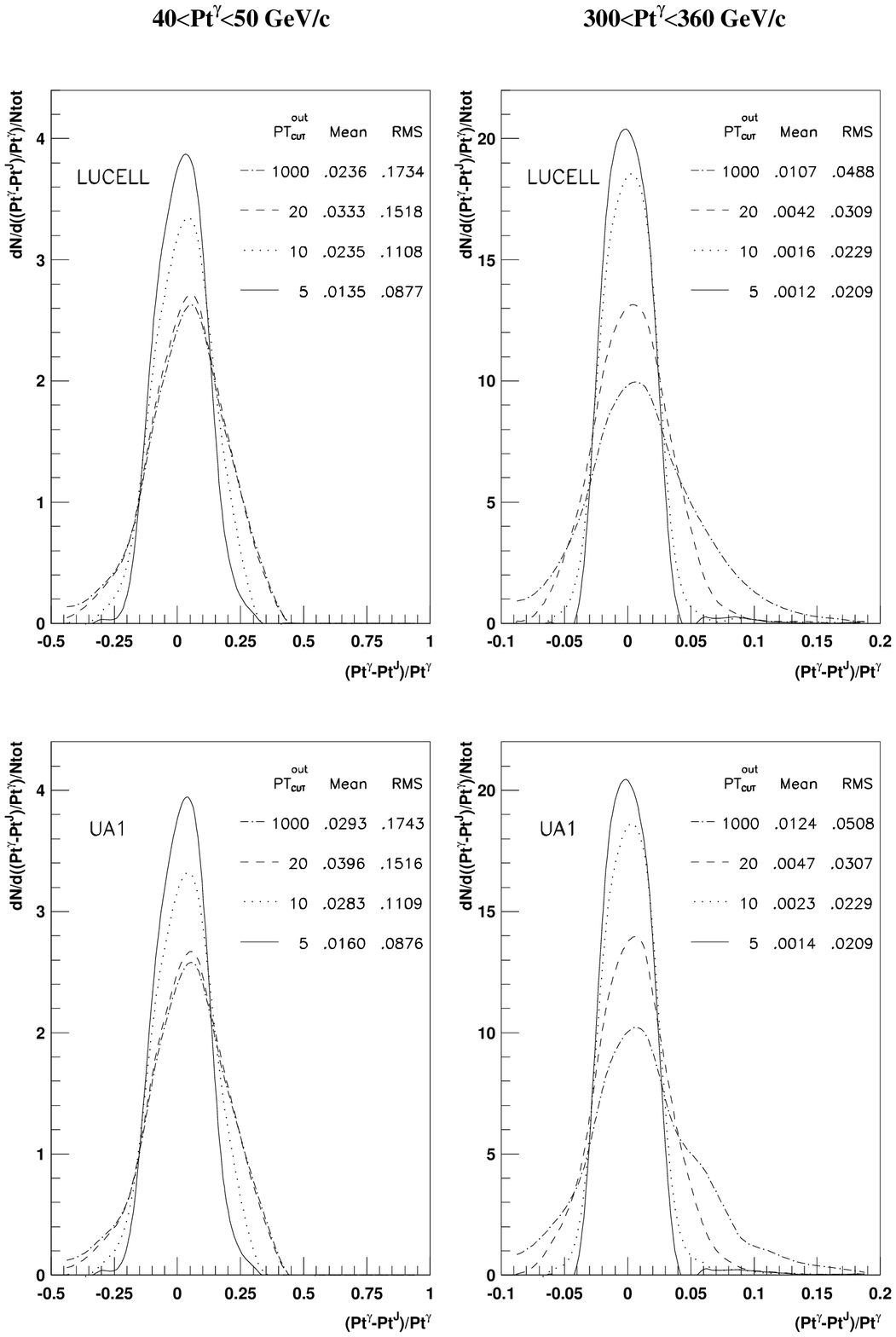}
   \caption{A dependence $(\Pt^{\gamma}-\Pt^{J})/\Pt^{\gamma}$ on
$\Pt^{out}_{CUT}$ for LUCELL and UA1 jetfinding algorithms and two
intervals of \ptg.  The mean and RMS of the distributions are displayed on
the figures. $\Pt^{clust}<30 ~GeV/c$. Selection 1}
\label{fig:j-out}
\end{figure}
\begin{figure}
  \hspace{0mm} \includegraphics[width=13cm]{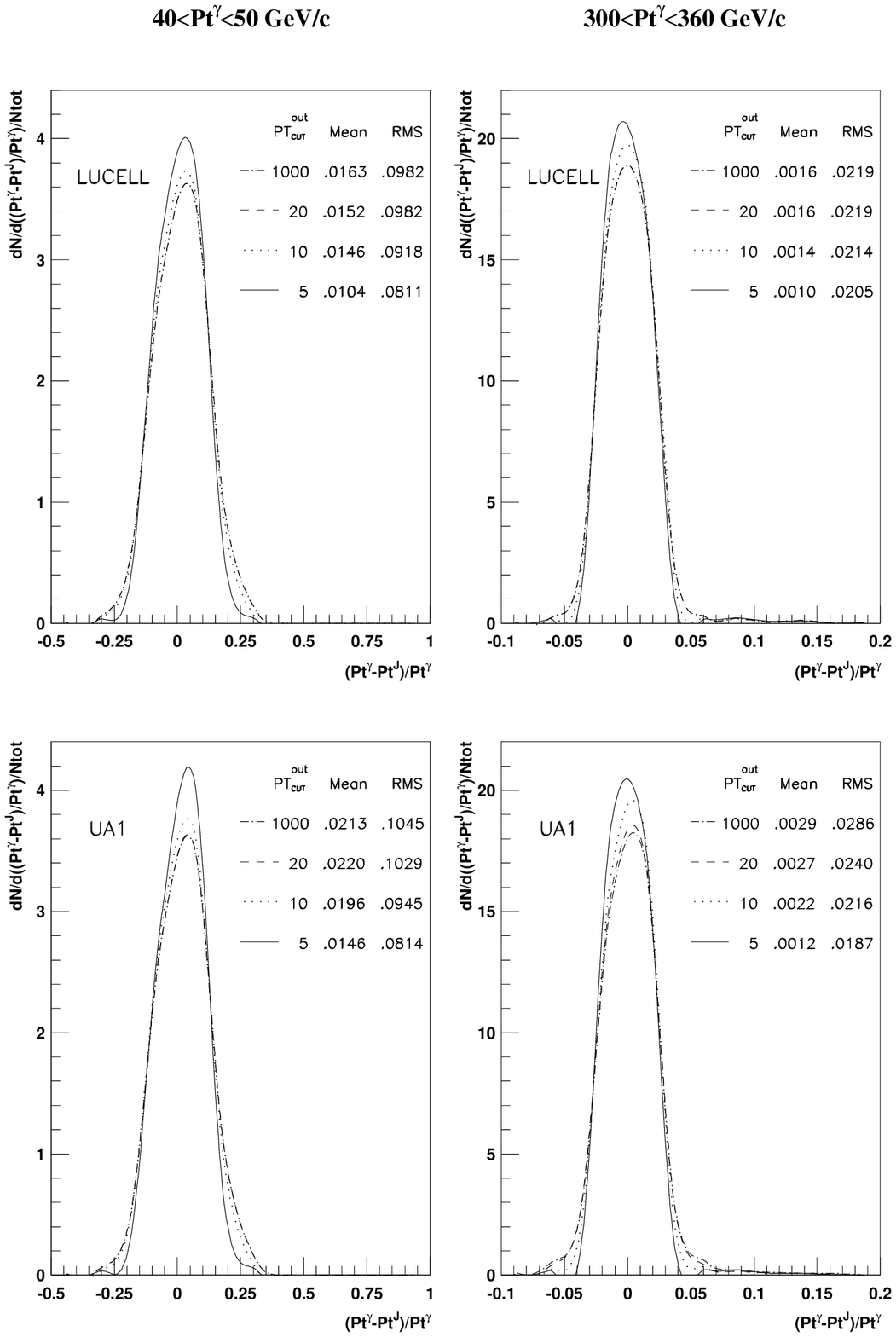}
   \caption{A dependence $(\Pt^{\gamma}-\Pt^{J})/\Pt^{\gamma}$ on
$\Pt^{out}_{CUT}$ for LUCELL and UA1  jetfinding algorithms and two
intervals of \ptg.  The mean and RMS of the distributions are displayed on
the figures. $\Pt^{clust}<10 ~GeV/c$. Selection 1}
\label{fig:j-out-}
\end{figure}

\end{document}